\documentclass[runningheads,a4paper]{llncs}
\usepackage{url}
\usepackage{hyperref}
\usepackage{multirow}
\usepackage{booktabs}
\usepackage{amsmath}
\usepackage{graphicx}

\usepackage{MnSymbol}

\urldef{\mailsa}\path|ameer.tawfik@gmail.com|
\urldef{\mailsb}\path|naomie@utm.my|   
\newcommand{\keywords}[1]{\par\addvspace\baselineskip
\noindent\keywordname\enspace\ignorespaces#1}
\begin{document}

\mainmatter
\title{Online Forum Thread Retrieval using Pseudo Cluster Selection and Voting Techniques}
\titlerunning{Forum Thread Retrieval using Pseudo Cluster Selection \& Voting Techniques}
\author{Ameer Tawfik Albaham 
		\and
        Naomie Salim
        }
        
\institute{Faculty of Computer Science and Information System,\\
Universiti Teknologi Malaysia, Skudai, Johor,Malaysia\\
\mailsa,
\mailsb
}

\maketitle
\begin{abstract}
Online forums facilitate knowledge seeking and sharing on the Web.
However, the shared knowledge is not fully utilized due to information overload. 
Thread retrieval is one method to overcome information overload.
In this paper, we propose a model that combines two existing approaches: the Pseudo Cluster Selection and the Voting Techniques. 
In both, a retrieval system first scores a list of messages and then
ranks threads by aggregating their scored messages. 
They differ on what and how to aggregate. 
The pseudo cluster selection focuses on input, while voting techniques focus on the aggregation method. 
Our combined models focus on the input and the aggregation methods. 
The result shows that some combined models are statistically superior to baseline methods.
\keywords{Forum thread search, Voting techniques}
\end{abstract}

\section{Introduction}
Online forums are user-generated content platforms, which enable users to build virtual communities. 
In these communities, users interact with each other to seek and share knowledge. 
The interaction happens through exchanging information in a form of discussions. 
A user starts a discussion by posting an initial message.
Then, replies to it are contributed by the other users. 
A pair of an initial message and the set of its reply messages builds a thread.

Thread retrieval helps forums' end users to find information satisfying their needs. 
However, the challenge is that threads are not text; they are collections of
"messages"---the initial and the reply messages. 
Therefore, given a user query, a retrieval system needs to utilize the text of the messages to rank threads. 
The problem of the thread retrieval is similar to the problem of the blog site retrieval\cite{seo2008,macdblog}. 
In blog site retrieval, given a query, we leverage the blogs' postings in order to rank blogs. 
An analogy between these two retrieval problems is that threads are the blogs, and messages are the postings.

Because of the thread retrieval's resemblance to the blog site retrieval, researchers, as in \cite{elsas2009} and \cite{seo2011}, have adapted techniques from blog distillation, such as \cite{elsas2008,seo2008}, to thread search; and, the adapted models performed well. 
Motivated by this fact, we propose to combine two known techniques on blog site retrieval:
Pseudo Cluster Selection(PCS)\cite{seo2008} and Voting Techniques(VT) \cite{macdblog}; 
then, we apply the combined model to thread search.

Both methods first rank a list of postings and then aggregate the postings' scores to rank their parent blogs. 
However, they differ in two aspects: the input and the aggregation method. 
PCS focuses on the top $k$ ranked postings from each blog, whereas VT considers all ranked postings. 
Furthermore, PCS uses the geometric mean to fuse scores, while VT adapts various data fusion methods, such as CombMAX\cite{df-fox} and expCombSUM\cite{ogilvie2003},
to the blog distillation task. In using CombMAX, blogs are ranked based on their best scoring posting;
in using expCombSUM, a blog is scored based on the sum of the exponential values of the blog's ranked postings' relevance scores.
It was reported that the voting methods that favor blogs with highly ranked postings performed better than the other voting methods\cite{macdblog}. 
This characteristic is similar to PCS's emphasis on highly ranked postings because it considers only the top $k$ postings.

In thread retrieval, \cite{elsas2011} reported that scoring threads using the maximum score
of their ranked messages is superior to using the arithmetic mean of the scores; and, 
PCS is statistically superior to both methods. 
Note that the arithmetic mean will be affected by the messages with low scores,
whereas the maximum score method favors threads with highly ranked messages. 
In addition, CombMAX is a special case of PCS($k$ = 1)\cite{elsas2009}. 
In other words, focusing on highly ranked messages improve the retrieval performance. 
Therefore, in this study, we hypothesize that voting methods can achieve better performance by focusing on only the top $k$ ranked messages.

\section{Related work}
The voting approach to the blog distillation task is inspired by works on data fusion\cite{df-fox,ogilvie2003} and expert finding \cite{macdexpert}. 
A data fusion technique aims to combine several ranked lists of documents generated by different retrieval methods into a unified list \cite{df-bordafuse-2001}.
In addition, retrieval methods that retrieved a document are voters for that document.
Then, a data fusion technique is used to fuse these votes.
Data fusion techniques are categorized into score based and rank based aggregation methods. 
The score based methods --- such as CombMAX\cite{df-fox}, use the relevance scores of documents,
whereas the rank based methods, such as BordaFuse \cite{df-bordafuse-2001}, 
utilize the ranking positions of these documents. 

The expert finding task is defined as retrieving a list of people who are experts in the topic of the user query\cite{macdexpert}.
To estimate the expertise of a person, methods in this task leverage the documents that are associated with or written by that person\cite{macdexpert}.
\cite{macdexpert} models the problem of expert finding as a data fusion problem.
Motivated by the success of the voting model in the expert finding task\cite{macdexpert}, 
\cite{macdblog} models the blog site retrieval as voting process as well.
The connection between these tasks is that: in both,
we first rank a list of documents with respect to a user query using an underlined text retrieval model,
then each ranked document is considered as a vote supporting the relevance of its associated "object"--- the person or the blog.
Indeed, in both tasks, the  voting approach was found to be statistically superior to baseline methods \cite{macdexpert,macdblog}.
Similarly, in the Pseudo Cluster Selection(PCS) method \cite{seo2008}, 
we first rank a list of postings, and then an aggregation method is applied. 
However, in PCS, only the top $k$ ranked postings from each blog are used to estimate the blog's relevance. 
If the number of the blog's ranked postings is less than $k$ , 
then a padding is supplied by using the minimum score of all ranked postings.

In thread retrieval, similar approach has been applied as well \cite{elsas2009,elsas2011,seo2011};
that is ranking threads by aggregating their message relevance scores. 
\cite{elsas2009} proposed two strategies to rank threads: inclusive and selective. 
The inclusive strategy utilizes evidences from all messages in order to rank parent threads.
Two models from previous work on blog site retrieval \cite{elsas2008} were adapted to thread search: the large document and the small document
models.
The large document model creates a virtual document for each thread by concatenating the thread's message text, then it scores threads based on their virtual document relevance to the query. 
In contrast, the small document model defines a thread as a collection of text units (messages). 
Then, it scores threads by adding up their messages' similarity scores.
In contrast to the inclusive strategy, \cite{elsas2009}'s selective strategy treats threads as a collection of messages; 
and it uses only few messages to rank threads. 
Three selective methods were used. 
The first one is scoring threads using only the initial message relevance score. 
The second method scores threads by taking the maximum score of their message relevance scores. 
The third method is based on PCS.
Generally, it was found that the selective models are statistically superior to the inclusive models \cite{elsas2009,elsas2011} especially the PCS method. 
Our work extends the Pseudo Cluster Selection method by investigating more aggregation methods.

Another line of research is the multiple context retrieval approach \cite{seo2011}. 
This approach treats a thread as a collection of several "local contexts" --- types of self-contained text units. 
Four contexts were proposed: posts, pairs, dialogues and the entire thread. 
The thread and post contexts are identical to \cite{elsas2009}'s virtual document and message concepts respectively. 
The pair and the dialogue contexts exploit the conversational relationship between messages to build text units. 
To rank threads using the post, pair and dialogue contexts, PCS was used. 
It was observed that retrieval using the dialogue context outperformed retrieval using other contexts. 
Additionally, the weighted product between the thread context and the post, pair or the dialogue contexts achieved better performance than using individual contexts. 
In our work, we are focusing on how to combine the ranked contexts' scores--- using the message context. 
Therefore, our work is complementary to \cite{elsas2009}'s work.

The third line of work is the structure based document retrieval \cite{bhatia2010}. 
In this approach, a thread consists of a collection of structural components: the title, the initial message and the reply messages set. 
In this representation, the thread relevance to the user query is estimated using \cite{metzler2004}'s inference network framework.
Our work can be applied to \cite{bhatia2010}'s representation as well. We could use \cite{bhatia2010}'s inference based relevance
score the same way the thread context score was used in \cite{seo2011}.

\section{Combined Models}
In this work, given a query $Q = \{q_1, q2,...,q_n\}$, 
we first rank a list of messages $R_Q$ with respect to $Q$. 
Then, we score threads by aggregating their top $k$ ranked messages' scores or ranks. 
\cite{seo2011} stated that the pair and the dialogue contexts must be extracted using thread discovery techniques, and retrieval using inaccurate extracted contexts will hurt the performance significantly. 
As a result, we use only the message context. 

In estimating the relevance between $Q$ and a message $M$, we employ the query language model assuming term independence,
uniform probability distribution for $M$ and Dirichlet smoothing\cite{zhai2004} as follows:
\begin{equation}
\label{eq-qlm}
  P(Q|M)=
  \prod_{q\in Q}
  	\left(
  		\frac
  		{n(q,M)+\mu P(q|C)}
  		{|M| + \mu}
  		\right)
  		^{n(q,Q)}
\end{equation}
where $q$ is a query term, $\mu$ is the smoothing parameter. $n(q,M)$ and $n(q,Q)$
are the term frequencies of $q$ in $M$ and $Q$ respectively, $|M|$ is the number of
tokens in $M$ and $P(q|C)$ is the collection language model.

To rank threads, the twelve aggregation methods proposed by \cite{macdexpert} are adapted:
Votes, Reciprocal Rank(RR), Bordafuse, CombMIN, CombMAX, CombMED,
CombSUM, CombANZ, CombMNZ, expCombSUM, expCombANZ and expCombMNZ. 
In addition to these methods, this study uses "CombGNZ"--- the geometric mean of the relevance scores. 
We use this method because it is the aggregation method employed by the Pseudo Cluster Selection method \cite{seo2008}.

Note that our combined models aggregate only the top $k$ ranked messages from each thread in order to infer a thread's relevance score. 
Let $R_T$ denote the set of all ranked messages from a thread $T$, and $R_{T,k}$ denote the set of the top $k$ ranked messages. 
However, if the size of $R_T$ is less than $k$, then the "set with
less members" is used--- either $R_T$ or $R_{T,k}$, we denote to it by $R_{T,L}$. 
Using $R_{T,L}$, we can score threads using any of the following methods:

\begin{equation}
\mbox{Votes}_k(Q,T) = |R_{T,L}|
\end{equation}

\begin{equation}
\mbox{RR}_k(Q,T) = \sum_{M \in R_{T,L}} \frac{1}{\mbox{rank}(Q,M)}
\end{equation}

\begin{equation}
\mbox{BordaFuse}_k(Q,T) = \sum_{M \in R_{T,L}} |R_Q| -− \mbox{rank}(Q,M)
\end{equation}

\begin{equation}
\mbox{CombMIN}_k(Q,T) = MIN_{M \in R_{T,L}} P(Q|M)
\end{equation}

\begin{equation}
\mbox{CombMED}_k(Q,T) = Median_{M \in R_{T,L}} P(Q|M)
\end{equation}

\begin{equation}
\mbox{CombSUM}_k(Q,T) = \sum_{M \in R_{T,L}} P(Q|M)
\end{equation}

\begin{equation}
\mbox{CombANZ}_k(Q,T) = \frac{1}{|R_{T,L}|} \times \sum_{M \in R_{T,L}} P(Q|M)
\end{equation}

\begin{equation}
\mbox{CombGNZ}_k(Q,T) = \left(\prod_{M \in R_{T,L}} P(Q|M) \right)^\frac{1}{|R_{T,L}|}
\end{equation}

\begin{equation}
\mbox{CombMNZ}_k(Q,T) = |R_{T,L}| \times \sum_{M \in R_{T,L}} P(Q|M)
\end{equation}

\begin{equation}
\mbox{expCombSUM}_k(Q,T) = \sum_{M \in R_{T,L}} \exp(P(Q|M))
\end{equation}

\begin{equation}
\mbox{expCombANZ}_k(Q,T) = \frac{1}{|R_{T,L}|} \times \sum_{M \in R_{T,L}} \exp(P(Q|M))
\end{equation}

\begin{equation}
\mbox{expCombMNZ}_k(Q,T) = |R_{T,L}| \times \sum_{M \in R_{T,L}} \exp(P(Q|M))
\end{equation}
where $\mbox{rank}(Q,M)$ is the rank of the message $M$ on $R_Q$, $|R_Q|$ is the size of
$R_Q$ and $|R_{T,L}|$ is the size of  $R_{T,L}$. 
This rest of this paper reports the retrieval performance using these methods.

\section{Experimental Design}
Thread retrieval is a new task, and the number of test collections is limited. 
In this study, we used the same corpus used by \cite{bhatia2010}. 
It has two datasets from two forums---Ubuntu\footnote{ubuntuforums.org} and
Travel\footnote{http://www.tripadvisor.com/ShowForum-g28953-i4-New York.html}
forums. 
The statistics of the corpus is as follows.
In the Ubuntu dataset, there are 113277 threads, 676777 messages, 25 queries and 4512 judged threads. 
In the Travel dataset,there are 83072 threads, 590021 messages, 25 queries and 4478 judged threads. 
The same relevance protocol was followed: a thread with 1 or 2 relevance judgement is considered as relevant, while a relevance of 0 is considered as irrelevant. 
Text was stemmed with the Porter stemmer and no stopword removal was applied. 
In conducting the experiments, we used the Indri retrieval system \footnote{http://www.lemurproject.org/indri.php}.

As for evaluation, we calculated Precision at 10
(P@10), Normalized Discounted Cumulative Gain at 10 (NDCG@10) and Mean Average Precision (MAP). 
In addition, we used the virtual document model($VD$)\cite{elsas2009} as our baseline. 
We used $VD$ because it has been used as a strong baseline in most previous studies\cite{elsas2009,seo2011}. 
In addition to $VD$, we used the basic aggregation methods.
In the basic methods, all ranked messages are included in the aggregation process. 
That enables us to make a fair judgement about the performance of the combined methods.

As for parameter estimation, we have three parameters to estimate: the smoothing parameters $\mu$ for the virtual document and message language models, the size of the initial ranked list of messages $R_Q$ and the value of $k$. 
To estimate $\mu$, we varied its value from 500 up to 4000; adding 500 in each run. 
To estimate the size of $R_Q$, we varied its value from 500 up to 5000 adding 500 in each run. 
To estimate $k$--- the number of top ranked messages, we varied its value from 2 up to 6 adding 1 in each run. 
Then, an exhaustive grid search was applied to maximize MAP using 5-fold cross validation.

\begin{table}
\begin{center}
\caption{Retrieval performance of fusion top $k$ messages on the Ubuntu dataset.}
\label{tbl-result-ra-u-top-k}

\begin{tabular}{lllll}
\toprule
 & Method      & MAP          & P@10 & NDCG@10\\
\midrule
Baseline & VD &  0.3437 $^{}$  &  0.4200 $^{}$  &  0.3284 $^{}$ \\

\midrule

\multirow{1}{*}{Basic models}
& Votes &  0.2749 $^{\triangledown}$  &  0.4680 $^{}$  &  0.3551 $^{}$ \\
& RR &  0.3313 $^{}$  &  0.4600 $^{\triangle}$  &  0.3428 $^{}$ \\ 
& BordaFuse &  0.3153 $^{}$  &  0.5080 $^{\triangle}$  &  0.3778 $^{}$ \\
& CombMIN &  0.1779 $^{\triangledown}$  &  0.2600 $^{\triangledown}$  &  0.1849 $^{\triangledown}$ \\
& CombMAX &  0.3074 $^{\triangledown}$  &  0.4480 $^{}$  &  0.3257 $^{}$ \\
& CombMED &  0.2212 $^{\triangledown}$  &  0.2760 $^{\triangledown}$  &  0.1927 $^{\triangledown}$ \\
& CombSUM &  0.3100 $^{}$  &  0.4720 $^{}$  &  0.3633 $^{}$ \\
& CombMNZ &  0.3108 $^{}$  &  0.4880 $^{}$  &  0.3720 $^{}$ \\
& CombANZ &  0.2314 $^{\triangledown}$  &  0.2800 $^{\triangledown}$  &  0.1991 $^{\triangledown}$ \\
& CombGNZ &  0.2272 $^{\triangledown}$  &  0.2760 $^{\triangledown}$  &  0.1971 $^{\triangledown}$ \\
& expCombSUM &  0.3088 $^{}$  &  0.4840 $^{}$  &  0.3676 $^{}$ \\
& expCombANZ &  0.2315 $^{\triangledown}$  &  0.2800 $^{\triangledown}$  &  0.1991 $^{\triangledown}$ \\
& expCombMNZ &  0.3088 $^{}$  &  0.4840 $^{}$  &  0.3676 $^{}$ \\

\midrule
\multirow{12}{*}{Top $k$ models}
& $\mbox{Votes}_k$ &  0.2699 $^{\triangledown}$  &  0.4600 $^{}$  &  0.3327 $^{}$ \\
& $\mbox{RR}_k$ &  0.3359 $^{}$  &  0.4600 $^{\triangle}$  &  0.3425 $^{}$ \\
& $\mbox{BordaFuse}_k$ &  0.3355 $^{\blacktriangle}$  &  0.4800 $^{}$  &  0.3659 $^{}$ \\
& $\mbox{CombMIN}_k$ &  0.2372 $^{\triangledown\blacktriangle}$  &  0.3320 $^{\triangledown\blacktriangle}$  &  0.2559 $^{\triangledown\blacktriangle}$ \\
& $\mbox{CombMED}_k$ &  0.2644 $^{\triangledown\blacktriangle}$  &  0.3680 $^{\blacktriangle}$  &  0.2698 $^{\triangledown\blacktriangle}$ \\
& $\mbox{CombSUM}_k$ &  0.3575 $^{\blacktriangle}$  &  0.4760 $^{\triangle}$  &  0.3851 $^{\triangle}$ \\
& $\mbox{CombMNZ}_k$ &  0.3544 $^{\blacktriangle}$  &  0.4800 $^{\triangle}$  &  0.3866 $^{\triangle}$ \\
& $\mbox{CombANZ}_k$ &  0.2644 $^{\triangledown\blacktriangle}$  &  0.3680 $^{\blacktriangle}$  &  0.2698 $^{\triangledown\blacktriangle}$ \\
& $\mbox{CombGNZ}_k$ &  0.2668 $^{\triangledown\blacktriangle}$  &  0.3840 $^{\blacktriangle}$  &  0.2832 $^{\blacktriangle}$ \\
& $\mbox{expCombSUM}_k$ &  0.3539 $^{\blacktriangle}$  &  0.4800 $^{\triangle}$  &  0.3865 $^{\triangle}$ \\
& $\mbox{expCombANZ}_k$ &  0.2644 $^{\triangledown\blacktriangle}$  &  0.3680 $^{\blacktriangle}$  &  0.2698 $^{\triangledown\blacktriangle}$ \\
& $\mbox{expCombMNZ}_k$ &  0.3539 $^{\blacktriangle}$  &  0.4800 $^{\triangle}$  &  0.3865 $^{\triangle}$ \\
\bottomrule
\end{tabular}
\end{center}
\end{table}

\section{Result and Discussion}
This section reports the result of fusing only the top $k$ messages. 
Table \ref{tbl-result-ra-u-top-k} and Table \ref{tbl-result-ra-t-top-k} presents the retrieval performance of these methods in the Ubuntu dataset and the Travel dataset respectively.
The symbols $^{\triangle}$ and  $^{\triangledown}$ denote statistically significant improvements or degradations over the virtual document model (VD) respectively,
whereas, $^{\blacktriangle}$ and  $^{\blacktriangledown}$ denote statistically significant improvements or degradations of  a top $k$ model over its basic model, e.g $\mbox{CombSUM}_k$ over CombSUM.
All significance tests are conducted using ttest at $p < 0.05$. 
The upper parts from the tables contain
retrieval performance of the virtual document (VD) and the basic aggregation methods, 
while the lower parts contain the performance of the combined methods.

Generally, the data on these tables supports the result from previous researches. 
In the basic mode, BordaFuse, CombSUM, CombMNZ, expCombSUM,
expCombSUM and expCombMNZ were able to produce better or comparable result with respect to $VD$. 
These methods favor threads with highly ranked messages. 
In contrast, CombGNZ, CombMED, CombANZ, CombMIN and expCombANZ might be affected by threads that have a lot of low scored messages.
In the combined models, all rank based methods have almost similar results to their results on the basic mode. 
In contrast, the score based methods benefit largely from fusing only the top k messages. 
Among them, methods favouring threads with highly ranked messages brought significant improvements over baseline methods--- see the performance of CombSUM, CombMNZ, expCombSUM and expCombMNZ. 
In addition, CombANZ, CombMIN, CombMED and expCombANZ benefit as well. 
Therefore, focusing on top $k$ messages is a good strategy to improve data fusion performance on thread retrieval.

\begin{table}
\begin{center}
\caption{Retrieval performance of fusion top $k$ messages on the Travel dataset.}
\label{tbl-result-ra-t-top-k}

\begin{tabular}{lllll}
\toprule
& Method & MAP & P@10 & NDCG@10 \\
\midrule
Baseline & VD &  0.3774 $^{}$  &   0.4800 $^{}$  &   0.3549 $^{}$ \\ 
\midrule
\multirow{12}{*}{Basic models}
& Votes &  0.2996 $^{\triangledown}$  &  0.5280 $^{}$  &  0.4190 $^{}$ \\
& RR &  0.3155 $^{\triangledown}$  &  0.4520 $^{}$  &  0.3432 $^{}$ \\
& BordaFuse &  0.3630 $^{}$  &  0.5640 $^{\triangle}$  &  0.4350 $^{\triangle}$ \\
& CombMIN &  0.1574 $^{\triangledown}$  &  0.3040 $^{\triangledown}$  &  0.2199 $^{\triangledown}$ \\
& CombMAX &  0.2727 $^{\triangledown}$  &  0.4360 $^{}$  &  0.3216 $^{}$ \\
& CombMED &  0.2004 $^{\triangledown}$  &  0.3480 $^{\triangledown}$  &  0.2389 $^{\triangledown}$ \\
& CombSUM &  0.3668 $^{}$  &  0.5560 $^{}$  &  0.4440 $^{\triangle}$ \\
& CombMNZ &  0.3575 $^{}$  &  0.5280 $^{}$  &  0.4205 $^{}$ \\
& CombANZ &  0.2065 $^{\triangledown}$  &  0.3400 $^{\triangledown}$  &  0.2346 $^{\triangledown}$ \\
& CombGNZ &  0.2000 $^{\triangledown}$  &  0.3320 $^{\triangledown}$  &  0.2319 $^{\triangledown}$ \\
& expCombSUM &  0.3513 $^{}$  &  0.5200 $^{}$  &  0.4109 $^{}$ \\
& expCombANZ &  0.2065 $^{\triangledown}$  &  0.3400 $^{\triangledown}$  &  0.2346 $^{\triangledown}$ \\
& expCombMNZ &  0.3513 $^{}$  &  0.5200 $^{}$  &  0.4109 $^{}$ \\
\midrule
\multirow{12}{*}{Top $k$ models}
& $\mbox{Votes}_k$ &  0.2864 $^{\triangledown\blacktriangledown}$  &  0.5280 $^{}$  &  0.3943 $^{}$ \\
& $\mbox{RR}_k$ &  0.3126 $^{\triangledown\blacktriangledown}$  &  0.4520 $^{}$  &  0.3400 $^{}$ \\
& $\mbox{BordaFuse}_k$ &  0.3639 $^{}$  &  0.5520 $^{}$  &  0.4380 $^{\triangle}$ \\
& $\mbox{CombMIN}_k$ &  0.1918 $^{\triangledown\blacktriangle}$  &  0.3360 $^{\triangledown\blacktriangle}$  &  0.2381 $^{\triangledown}$ \\
& $\mbox{CombMED}_k$ &  0.2478 $^{\triangledown\blacktriangle}$  &  0.3960 $^{\triangledown}$  &  0.2881 $^{\triangledown}$ \\
& $\mbox{CombSUM}_k$ &  0.3736 $^{}$  &  0.6080 $^{\triangle\blacktriangle}$  &  0.4788 $^{\triangle}$ \\
& $\mbox{CombMNZ}_k$ &  0.3608 $^{}$  &  0.5720 $^{\triangle}$  &  0.4587 $^{\triangle}$ \\
& $\mbox{CombANZ}_k$ &  0.2478 $^{\triangledown\blacktriangle}$  &  0.3960 $^{\triangledown}$  &  0.2881 $^{\triangledown\blacktriangle}$ \\
& $\mbox{CombGNZ}_k$ &  0.2330 $^{\triangledown\blacktriangle}$  &  0.3760 $^{\triangledown}$  &  0.2754 $^{\triangledown}$ \\
& $\mbox{expCombSUM}_k$ &  0.3600 $^{}$  &  0.5720 $^{\triangle}$  &  0.4559 $^{\triangle}$ \\
& $\mbox{expCombANZ}_k$ &  0.2479 $^{\triangledown\blacktriangle}$  &  0.3960 $^{\triangledown}$  &  0.2882 $^{\triangledown\blacktriangle}$ \\
& $\mbox{expCombMNZ}_k$ &  0.3600 $^{}$  &  0.5720 $^{\triangle}$  &  0.4559 $^{\triangle}$ \\
\bottomrule
\end{tabular}
\end{center}
\end{table}

\subsection*{Discussion}
The improvement might be due to the removal of noise introduced when all ranked messages are considered. 
By focusing on the top $k$ messages, low score messages are discarded leading to improving retrieval. 
In fact, ranking using only one message, e.g. CombMAX, might not be enough to capture the topical
relevance of threads, while considering all ranked messages might introduce irrelevant messages. 
Therefore, focusing on the top $k$ ranked messages will balance these two aspects. 
This premise explains, to some extend, why no improvements were observed among rank based methods.
Rank based methods inherently address the issue of low scoring messages.
Furthermore, the near similar performance of CombSUM, CombMNZ and their exponential variants suggests that:
once we focus on the top $k$ messages, extra emphasis on these messages does not help.
In other words, focusing on the top ranked messages is the main reason behind the observed improvements.

\begin{figure}
   \centering
   \includegraphics[width=0.7\textwidth]{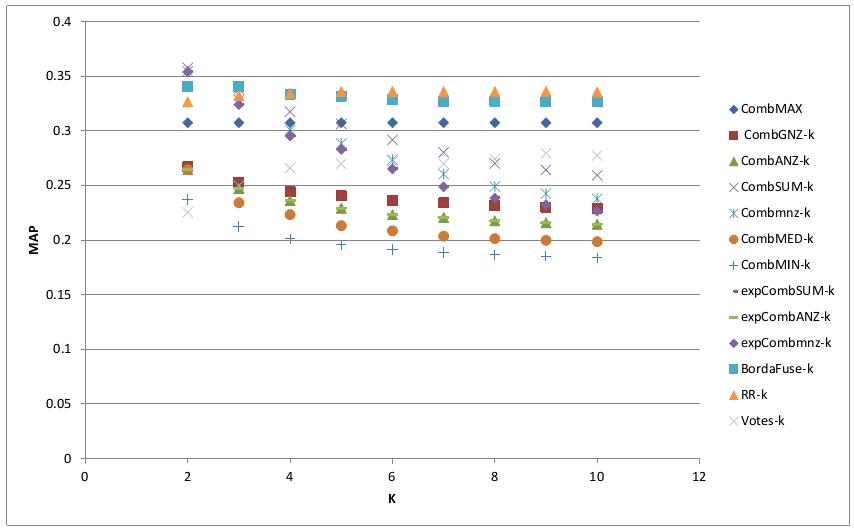} 
   \caption{The performance of the combined methods as $k$ changes on the Ubuntu dataset.}
  \label{fig-u-ra-navie-size-map}
          
\end{figure}

\begin{figure}
  \centering
   \includegraphics[width=0.7\textwidth]{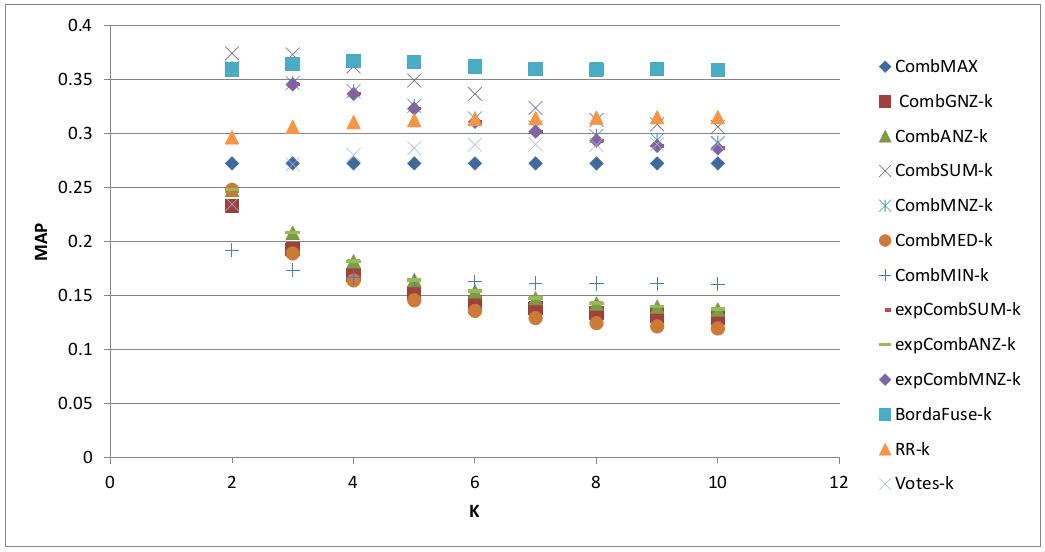}
  \caption{The performance of the combined methods as $k$ changes on the Travel dataset.}
  \label{fig-t-ra-navie-size-map}       
\end{figure}

To confirm this hypothesis, a study was conducted to investigate how does the performance change as $k$ increases? 
As Figures \ref{fig-u-ra-navie-size-map} and \ref{fig-t-ra-navie-size-map} show, the optimal value of $k$ for
each method ranges between 2 to 5 messages. 
Going beyond the optimal value, 
rank based methods tend to have consistent performances that are similar to their result on the basic mode. 
In contrast, score based methods performed badly as $k$ increases. 
That further supports the proposed hypothesis.
Another interesting observation is the performance of CombGNZ.
The result shows the inferiority of CombGNZ to the aforementioned good performing methods. 
However, \cite{elsas2009,elsas2011} reported that PCS--- using the geometric mean, outperformed CombMAX. 
That does not contradict our findings. In \cite{elsas2009,elsas2011}, PCS
adds a padding step if the number of ranked messages is less than $k$, whereas we
do not apply this step. We plan to study the effect of the padding step in future works.

\section{Conclusion and Future Works}
In this paper, we addressed the problem of online forums thread retrieval.
We conducted experiments to investigate the performance of combining the pseudo cluster selection and voting techniques approaches.
The results showed that the combined models can improve the performance of score based voting techniques significantly in various measures.

Our future work has two directions. 
First, we will apply the combined methods as aggregation methods on the other thread representations such as fusing the dialogue or the pair relevance scores \cite{seo2011}. 
Second, we will study the effect of the padding step on the performance of the combined models.

\subsubsection*{Acknowledgements} 
This work is supported by Ministry of Higher Education (MOHE) and 
Research Management Centre (RMC) at the Universiti Teknologi Malaysia (UTM) under Research University Grant Category ( VOT Q. J130000 .2528 .02H99 ).

\end{document}